# Title: Machine learning-driven complex models for wavefront shaping through multimode fibers

**Authors:** Jérémy Saucourt[1], Benjamin Gobé[1], David Helbert[2], Agnès Desfarges-Berthelemot[1], and Vincent Kermene[1*]

[1]*XLIM Research Institute, UMR 7252 CNRS/University of Limoges, Limoges F-87000, France*
[2]*XLIM Research Institute, UMR 7252 CNRS/University of Poitiers, Poitiers F-86000, France*
*[vincent.kermene@xlim.fr](mailto:vincent.kermene@xlim.fr)

**Abstract:** We investigate a method to retrieve full-complex models (Transmission Matrix and Neural Network) of a highly multimode fiber (140 LP modes/polarization) using a straightforward machine learning approach, without the need of a reference beam. The models are first validated by the high fidelity between the predicted and the experimental images in the near field and far field output planes (Pearson correlation coefficient between 97.5% and 99.1% with our trained Transmission Matrix or Neural Network). Their accuracy was further confirmed by successful 3D beam shaping, a task achievable only with a true full complex model. As a prospect, we also demonstrate the ability of our neural network architecture to model nonlinear Kerr propagation in gradient index multimode fiber and predict the output beam shape.

1. Introduction

Over the past decade, numerous studies have demonstrated that multimode fibers (MMF) with single or multiple cores hold significant potential for a wide range of applications including optical telecommunications [1-3], bio-imaging [4-6], fiber-based sensors [7] high power lasers or delivery systems [8,9] and optical manipulation [10,11]. These applications take advantage of the multiple modes and their interactions as degrees of freedom that can be harnessed within a single fiber. At low power, the MMF performs a deterministic linear transformation which depends on the opto-geometrical characteristics of the fiber, its packaging and the launching conditions of the input field into the MMF. Consequently, any coherent beam injected into an MMF produces a complex random pattern (speckle) very different from that at the fiber input. The induced aberrations can be typically controlled by a spatial light modulator (SLM) which pre-compensates the wavefront of the input beam [12]. The linear transformation effected by the MMF which accounts for the propagation and the interactions between the guided modes can be simply modelled by a transmission matrix ($TM$). This matrix links the complex field at the fiber input to the field at its output, thereby describing the intermodal coupling [13]. Many studies and applications exploiting an MMF rely on this model, particularly for image transmission or projection [14-15], and optical communication channels [16]. The full complex TM of an MMF is typically measured using off-axis holography with a reference beam [14], which adds complexity to the setup, especially for characterization of a long MMF. Recently, a new paradigm has emerged for characterizing the $TM$ of an MMF without a reference beam, using machine learning. Instead of measuring the relationship between coherent fields coupled into the fiber and the resulting output speckle intensity patterns, the methods learn this relationship directly from datasets made of intensity patterns from the MMF, and their corresponding input fields shaped by an SLM. Various optimization processes have been developed to solve the non-linear equations relating these data pairs, and retrieve the $TM$. Methods such as gradient descent [17], alternating projection [18], Bayesian inference [19,20] and semi-definite programming [21] have been used to obtain the $TM$ of an MMF without the need of a reference beam. However, because these processes compute the $TM$ row by row independently, without any phase constraint, there remains an unknown phase bias between its rows. As a result, these $TM$ can only be used to optimize the shape of the MMF output beam and do not provide any information on mode coupling for instance. Very recently, V. Tran *et al.* proposed to eliminate the phase bias of the $TM$ by measuring phase shifts between experimentally established focal points. However, this method is prone to error diffusion between the rows of the matrix, which can have detrimental effects on the usability of the $TM$ [22]. These last years, artificial neural networks ($NN$) have emerged as a promising alternative to the $TM$ approach, particularly because these digital twins have demonstrated resilience to environmental perturbations [23] and are well-suited to describe nonlinear transformations [24]. They have been applied to three main types of application: image transmission, image projection and mode decomposition. In the first two cases, the $NN$ learns the nonlinear backward mapping relationship between the MMF output speckle and the corresponding image displayed at the fiber input. The $NN$ is used either to retrieve an MMF input information from the scrambled output beam [25,26] or to predict the input wavefront that will produce a targeted image at the MMF output [27]. It is worth noting that for image projection, in [25], the authors needed to learn two models: one for the backward mapping (from MMF output intensity to input wavefront shape) and one for forward propagation (from SLM to MMF output). Additionally, $NN$ can also decompose the output speckle of an MMF into its modal basis, to retrieve the corresponding complex field [28]. However, these modal decompositions by $NN$s are limited to a small number of modes. Whatever the digital twin ($TM$ or $NN$) learnt by machine learning, without interferometer setup and reference beam, these models are most of the time not fully complex, i.e. does not link the input and output complex fields and consequently cannot predict the complex output field from the input complex field. In what follows, we report on original learning processes providing full complex digital twins of both $TM$ and $NN$ types. Our objective was to develop processes which can be implemented with a simple setup only based on a SLM at the MMF input and no more than standard two intensity detections at the output. We validated these models by successfully achieving 3D beam shaping through a highly multimode fiber and examined their inference capability, specially their capability to produce target shapes that were not induced during the training phase.

2. **Data set and setup for training the MMF full complex models**

The models we developed, link the input field $x \in \mathbb{C}^{n_x \times n_x}$ structured by the SLM, and the output field $y \in \mathbb{C}^{m_y \times m_y}$ (see Fig. 1). Without any reference beam, these models are typically trained using a set of data pairs $[X, |Y|^2]$, where $X \in \mathbb{C}^{n_x \times n_x \times N}$ and $|Y|^2 \in \mathbb{R}_*^{+^{m_y \times m_y \times N}}$, with $N$ being the number of data pairs. Numerical simulations have shown that the number of data pairs $N$ depends not only on the model architecture, but also on the number of phase modulating pixels on the SLM. For a TM-type model, $N > 20n_x^2$, whereas for a NN-type model, $N > 40n_x^2$. The set of intensity speckles $|Y|^2$ are recorded by a camera over a region of interest of $m_y \times m_y$ pixels, generated by the corresponding set of data $X$ randomly structured by an SLM. $TM$ or $NN$ learnt with these pairs of data, produce output intensity patterns $|y|^2$ close to the ground truth. Actually, they compute a complex field with a random phase bias between each sample of the field, because no phase constraint is applied on the data $|y|^2$ that feed these models. To obtain the full complex models, we added measurements $|z|^2 \in \mathbb{R}_*^{+^{m_z \times m_z}}$ in the far field of the MMF output beam to eliminate the phase bias.

We tested the two-plane measurement process with the experimental setup shown in Fig. 1. This setup included a 50μm-core, 0.22 numerical aperture MMF (step index), guiding up to 140 LP modes per polarization at 1064nm. The laser source was a CW DFB laser diode (QLD1061 from QDLaser), linearly polarized, delivering a fundamental Gaussian beam. It illuminated an SLM of segmented deformable mirror type with 952 independent actuators adjusting piston phase only (round shape, 34 actuators/diameter, Kilo-CS-0.6-SLM from Boston Micromachines Corporation). The SLM plane was imaged on the input fiber facet by a couple of afocal optical systems and the MMF output near and far fields were imaged on two 16-bit CMOS cameras (ThorCam CS2100). We considered a single linear polarization state at the fiber output.

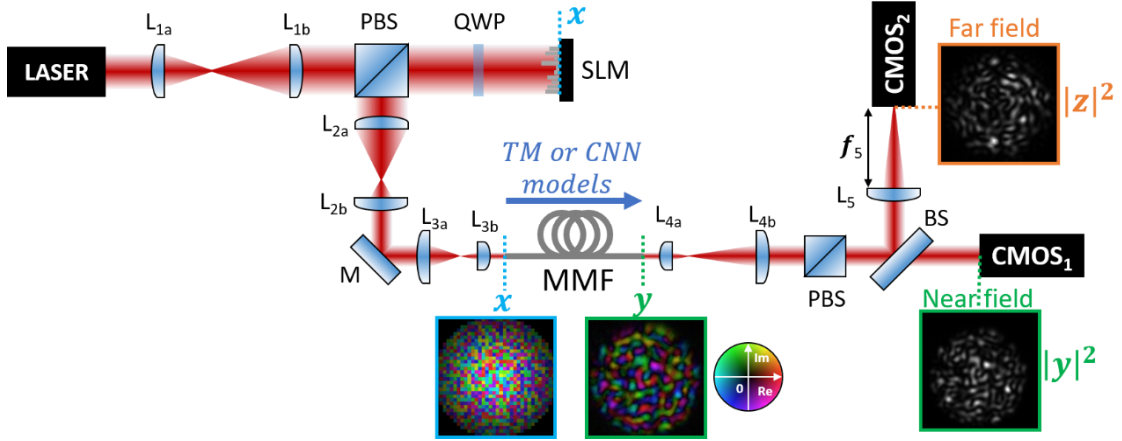

*Figure 1: Experimental setup designed for learning TM or NN models and beam shaping. Lia-Lib: imaging system "i"; PBS: Polarizing Beam Splitter; QWP: Quarter wave plate; SLM: Spatial Light Modulator; CMOSi: CMOS camera "i"; M: plane mirror; MMF: Multimode fiber; BS: Beam Splitter; $L_{ia,b}$: lens of the ith afocal.*

### 3. Full-complex Transmission Matrix model of the MMF computationally estimated by machine learning

For $TM$ learning, the 2D data $(x, |y|^2, |z|^2)$ are reduced to vectors of respective dimensions $n = n_x^2$, $m = m_y^2$ and $m$ The first equation to solve is of the form:

$$|y|^2 = |A.x|^2 \qquad (1)$$

where $(A \in \mathbb{C}^{m \times n})$ is the $TM$ to retrieve. This equation has got an infinite number of solutions as any matrix $\hat{A}$ verifies:

$$\hat{A} = Diag(\exp(j\psi)).A \qquad (2)$$

($Diag$ a diagonal matrix and $\psi \in \mathbb{R}^m$ a random phase bias between each line of the matrix $A$.

We first optimized the $TM$ coefficients of the MMF with a mini-batch gradient descent algorithm described in the reference [29] minimizing the loss function $\mathcal{L}_1$:

$$\min_{A \in \mathbb{C}^{m \times n}} \mathcal{L}_1 = \||Y|^2 - |A.X|^2\|^2 \qquad (3)$$

The dataset consisted of $Nc$ phase maps and their related near field intensity patterns at the MMF output, 80% of this dataset was used for training the $TM$ while the remaining 20% was allocated for validation. All data are recorded in a single experimental session, then randomly shuffled and divided in both groups of training and testing sets. We numerically demonstrated that an efficient model of the $TM$ requires a dataset size $N > 9n_x^2$. Results shown in the following were obtained with $N =$

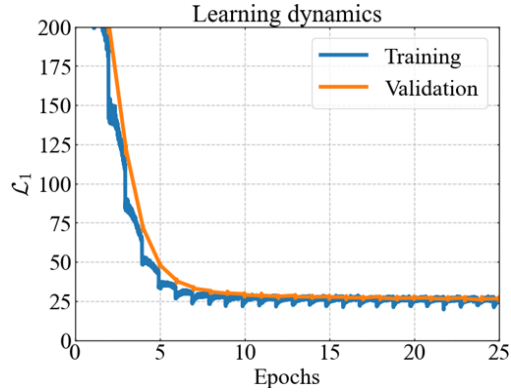

Figure 3 : Learning dynamics of the MMF TM

$17340 \sim 18n_x^2$ ($n_x^2 = 952\ SLM\ actuators$). Figure 2 shows the learning dynamics that converges in less than 25 epochs of 32-mini batch size. Figure 3 compares, for the same example of input field (identical phase map on the SLM), the output speckles computed with the retrieved $TM$ $\hat{A}$ and the ones measured by the camera on both output planes (near field and far field). As expected, the intensity pattern in the near field obtained with $\hat{A}$ closely matches the actual pattern (Pearson correlation coefficient $\Gamma(|y|^2, |\hat{A}.x|^2)$ = 99% between the ground truth $|y|^2$ and $|\hat{A}.x|^2$ computed with the $TM$ $\hat{A}$ (see Fig. 3), while the beams in the far field do not. As illustrated in Fig.3, the phase in the near field is pixel to pixel randomly distributed.

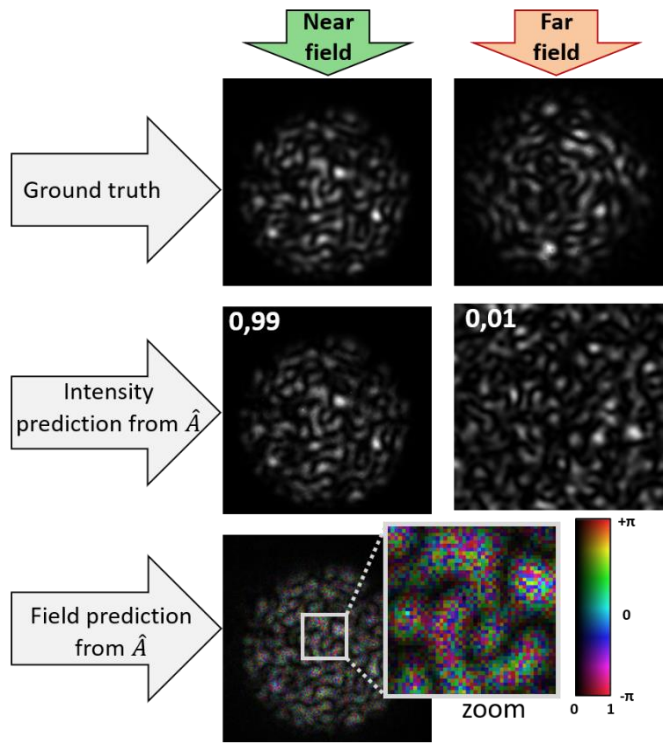

Figure 2 : Example of ground truth (experimental) and computed output beam with $\hat{A}$ in near field and far field (normalized gray scale in intensity). Zoom of the phase structure given by $\hat{A}$ (complex amplitude colorbar). Pearson correlation coefficients are indicated for intensity prediction from $\hat{A}$.

We corrected the phase biases $\psi$ by recording $k = 20$ additional images in the Fourier plane to minimize the loss function $\mathcal{L}_2$:

$$\min_{\psi \in [-\pi, +\pi[^n} \mathcal{L}_2 = \sum_{i=1}^{k}\left(1 - \Gamma\left(\left|FT(\hat{A}.x_i)\right|, |z_i|\right)\right) \quad (4)$$

$FT$ the 2D fast Fourier transform. The phase bias $\psi$ is retrieved by gradient descent algorithm [29] in less than 500 iterations. The new retrieved full complex TM can be rewritten as:

$$A_f = Diag(\exp(-j\psi)).\hat{A}. \quad (5)$$

This full complex $TM$ can now compute both near and far fields with a high accuracy as demonstrated in Fig.4.

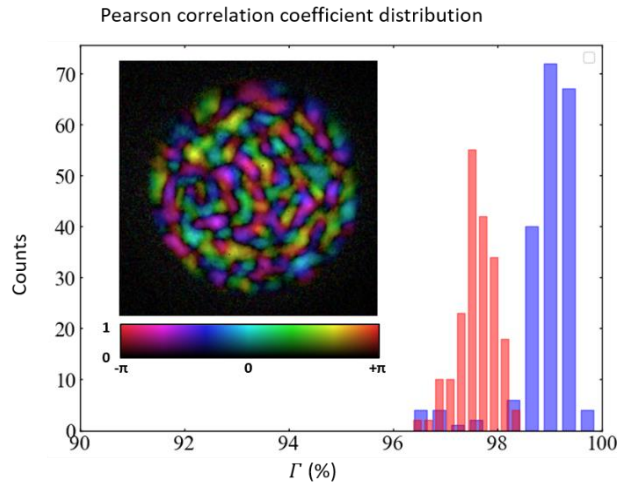

*Figure 4 : Distribution of Pearson correlation coefficients between the computed speckles and the corresponding ground truth patterns in the near field ($\Gamma(|y|^2, |A_f.x|^2$, blue bars) and far field ($\Gamma(|z|^2, |FT(A_f.x)|^2$, red bars) from the validation set of data.*

This figure shows statistical similarities between ground truths and computed beams in both planes, with a very high Pearson correlation coefficient (medians: 99,1% in near field and 97.6% in far field). An example of complex field obtained with the $A_f$ matrix is shown as an inset in Fig. 4. Contrary to the image of the Fig. 3, each speckle grain exhibits a more realistic smooth phase. We transformed the $A_f$ matrix to visualize the complex field generated by each actuator of the SLM through the MMF. Each column of $A_f$ was reshaped as an image and positioned to the area of the related actuator (Fig. 5) displaying the SLM as a combination of these complex fields. We can note that some actuators of the SLM are inactive and appear dark on this figure. As expected, the central actuators more likely excited low-order modes (Fig.5 upper right) while the peripheral actuators excited high-order modes (Fig. 5 lower right). To further validate the full complex matrix beyond comparing computed and measured beams in both near field and far field, we used $A_f$ to shape the MMF output beam in additional planes. This was first achieved by selecting a Rayleigh plane to place the target shape, which can be of complex value, then back-propagating this field to the output MMF facet and multiplying the resulting field by the inverse $TM$. However, since we can only adjust the phase of the input field, this method results in a poor similarity between the target field and the one produced by the model using phase-only modulation. This is why we finally used $A_f$ to optimize the SLM phase map and to achieve a shaped beam as close as possible to the target field. To accomplish this, $A_f$ needs to be refined by removing its unrealistic singular values and prevent non-physical beam shape optimization with excessively high resolution. First, we set to zero the normalized singular values (obtained through Singular Value Decomposition) whose indexes are higher than the number of modes (280 for both polarizations) guided by the MMF, thus minimizing noise channels (Fig. 6).

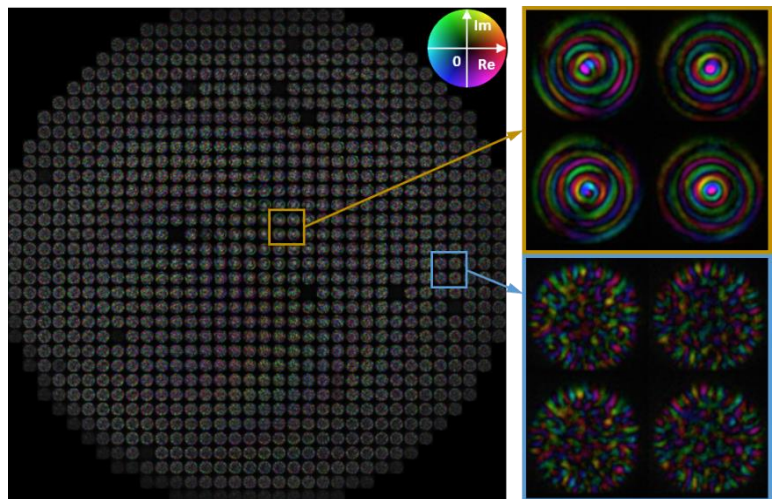

*Figure 5 Left: reshaped $A_f$, mapping each column of the learned matrix $A_f$ as a complex field and assigning it to the position of the related actuator on the SLM (34x34 actuators, ROI of the camera: 256x256 pixels) - Upper right: zoom on 4 fields from the central actuators (4 columns of $A_f$ related to 4 actuators from the central part of the SLM) - Lower right: zoom on 4 fields from the peripheral actuators (4 columns of $A_f$ related to 4 actuators from the peripheral part of the SLM).*

We then used this filtered transmission matrix for 3D beam shaping. As a demonstration, we simultaneously optimized two different beam shapes in two different planes, outside those where the learning data for $A_f$ were collected at the MMF output. In the first plane, $L_1 = 35\mu m$ from the fiber output, we selected a digit as target shape, while in the second plane, $L_2 = 70\mu m$ from the fiber output, we chose a letter. We minimized the loss function $\mathcal{L}_3$ to determine the phase map $\phi$ that would generate the target shapes in both planes simultaneously:

$$\min_{\phi \in [-\pi, +\pi[^n} \mathcal{L}_3 = \sum_{i=1}^{2} \left(1 - \Gamma\left(\left|\mathcal{F}r_{L_i}(A_f \cdot |x|e^{j\phi})\right|, |z_{L_i}|\right)\right) \qquad (6)$$

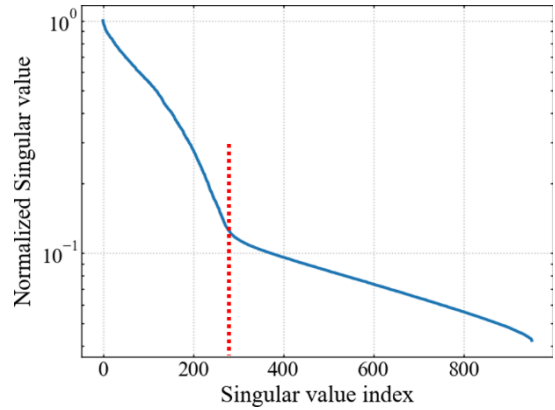

Figure 6 : Normalized singular values of the TM $A_f$. Singular values of index larger than the number (red dot line) of the MMF LP modes (both polarizations) were filtered before using the TM to compute the target phase map for beam shaping

, where $\mathcal{F}r_{L_i}$ is the Fresnel transform to reach the plane $L_i$ from the MMF output. The couples of beam shapes ([1,a], [2,b] .. [9,i]) obtained experimentally exhibit a good similarity with those computed using $A_f$ (Fig.7), providing another validation of the full complex model of the MMF measured without any reference beam. This experiment also demonstrates the ability of the process to perform 3D beam shaping, which is not feasible with a standard phase biased TM retrieved by machine learning.

It is worth noting that we retrieved a pixel-basis $TM$, from actuators of the SLM to pixels of the camera. Since this $TM$ is full complex, it can be converted to a modal basis. This conversion requires transformations at both side of the fiber, that consider the projection of the beam, through the optical imaging systems and their aberrations, to the modal basis of the MMF [30]. While this additional operation is beyond the scope of this article, it may be valuable for certain applications that involve analyzing modal coupling in the fiber.

In addition to traditional transfer matrices, several digital models of MMFs have been developed using neural networks of various architectures. However, to date, these $NN$ models in the MMF context, designed for tasks like image transmission or image projection through MMFs, are rarely providing

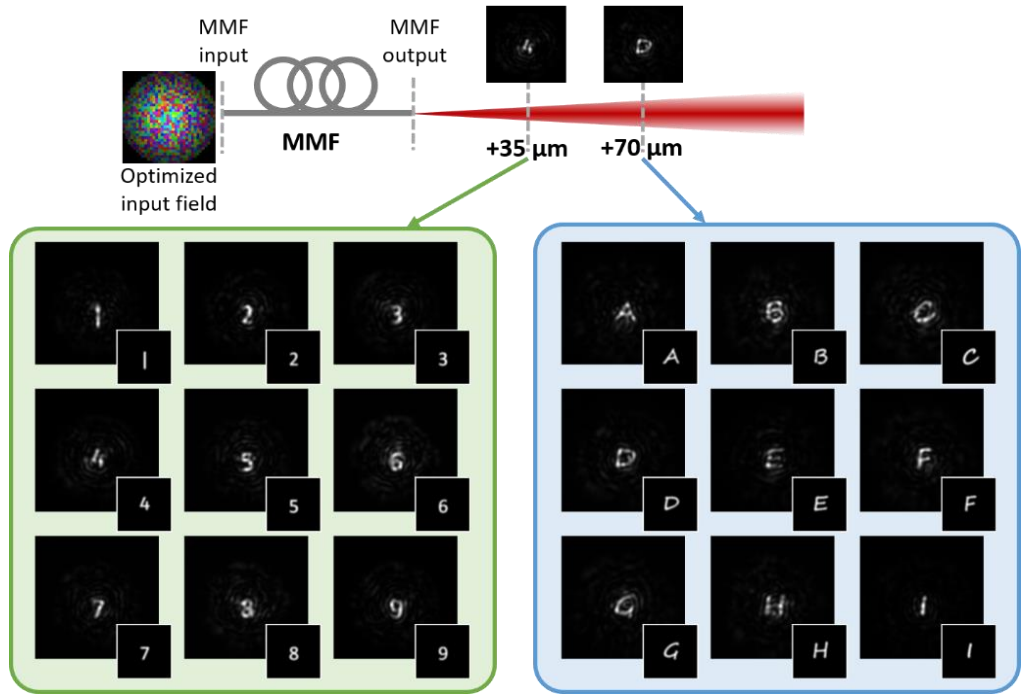

*Figure 7: Examples of shaped beams at the MMF output in two distinct planes simultaneously, arbitrary chosen at 35µm and 70 µm from the fiber exit, to be compared with the target shapes (inset)*

with complex field predictions. In the following section, we detail our dedicated work on training $NNs$ specifically aimed at predicting complex fields. This work paves the way for generalizing MMF models to manage 3D propagation more effectively.

4. **Convolutional Neural Network as a full-complex MMF model**

We extracted the Generator $NN$ of the conditional Generative Adversarial Neural network (cGAN) used in the reference [31] and adapted it to our complex images. This Generator is a Convolutive Neural Network ($CNN$) that transforms images made of 3 channels dedicated to the red, green and blue colors into other RGB images. We modified this configuration to exploit these channels for complex data. This $CNN$ is a U-Net network learned to predict the complex field $y \in \mathbb{C}^{n \times n}$ at the fiber output ($n \times n$ samples) from a complex field $x \in \mathbb{C}^{n \times n}$ at the fiber input: $y = CNN(x)$. The complex field $x$ was decomposed in modulus, sine and cosine functions, and used as the 3 $CNN$ input channels $[|x|, \cos(\arg(x)), \sin(\arg(x))]$. Two of the three output channels of the $CNN$ are selected as the real and imaginary parts $[Real(y), Im(y)]$ of the output complex field (the remaining channel is unused). In our data management process, the neural network handles only real data (images of size $nxnx3$), and outputs the real and imaginary parts of the complex data used to reconstruct the complex field. The depth of the initial $CNN$ was reduced to 5 down-sampling and up-sampling blocks to consider a 64x64 complex images of 12-bit discretization (camera bit-depth). The architecture of the $CNN$, detailed in [32], is depicted on Figure 8, showing the size and composition of the different blocks. In an original way, this $CNN$ is trained using triplets of data $[x, |y|^2, |z|^2]$. They correspond respectively to the known complex fields at the fiber input, structured by the SLM, and to the intensity images in the near field and far field at the fiber output recorded by a couple of cameras. This setup is identical to the one used to measure the $TM$ (Fig. 1). Unlike the learning process for retrieving the full complex $TM$, the $CNN$ was trained in a single step. However, it requires knowledge of the corresponding far field intensity image for each near field intensity. The global loss function $\mathcal{L}_{CNN}$ was minimized using the ADAM (ADAptative Moment estimation) optimizer, considering the experimental set of images from both near field and far field planes as well as the numerical aperture ($NA$) of the MMF (Fig. 9). $\mathcal{L}_{CNN}$ is defined as:

$$\mathcal{L}_{CNN} = 3 - \left[\Gamma(|CNN(x)|^2, |y|^2) + \alpha \, SSIM\left(\left|FT(CNN(x))\right|^2, |z|^2\right) + PIB_{NA}\left(\left|FT(CNN(x))\right|^2\right)\right] \quad (7)$$

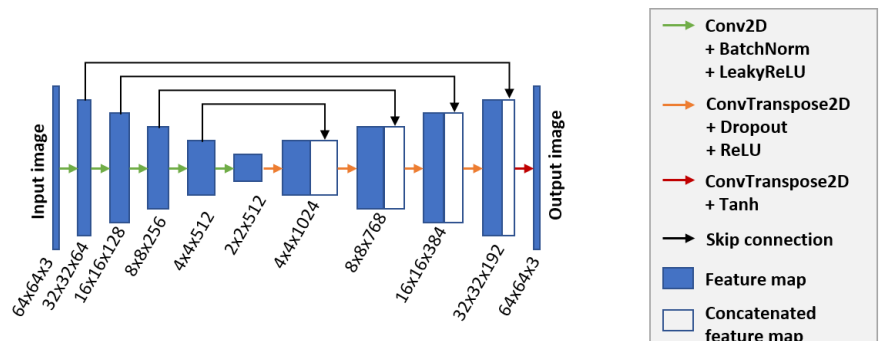

*Figure 8 : 3-channel CNN architecture (U-NET)*

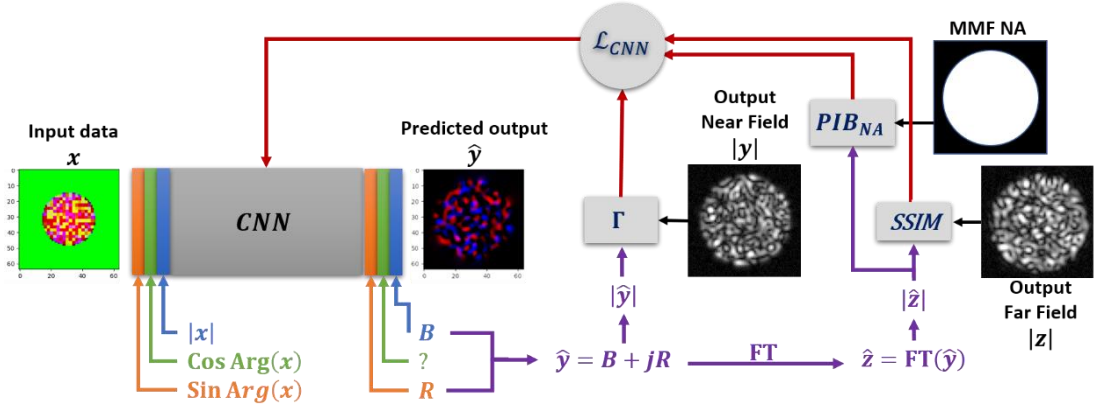

Figure 9: CNN learning process highlighting the triplet of data in the loss function $\mathcal{L}_{CNN}$ of the U-Net-type convolutional neural network computed to model the MMF. $\mathcal{L}_{CNN}$ is computed using three parameters: $\Gamma$ (Pearson Correlation Coefficient), SSIM (Structural Similarity Index Measurement) and $PIB_{NA}$ (Power ratio in the bucket, i.e contained in the MMF numerical aperture). The output "green" channel with a "question mark" is not used to train the CNN and can be removed without consequence on the results.

$\Gamma$ is the Pearson Correlation Coefficient between measured $|y|^2$ and predicted $|CNN(x)|^2$ patterns, $SSIM$ is the Structural Similarity Index Measurement between measured $|z|^2$ and predicted $\left|FT(CNN(x))\right|^2$ intensities. $PIB_{NA}$ is the ratio of the predicted energy within the numerical aperture of the fiber over the whole computed energy $\left|FT(CNN(x))\right|^2$. $\mathcal{L}_{CNN}$ is an adaptive loss function weighted by the factor $\alpha \in [0, 1]$, as explained in [32]. At the start of the CNN training process, $\alpha$ was close to 0 and gradually increased to 1 as $\Gamma(|CNN(x)|^2, |y|^2)$ approached 1. This is a quasi-sequential process that significantly enhances the convergence speed of the optimizer. Initially, it focuses on intensity prediction in the near field, and then progressively refines the phase content by improving the far field prediction. In the training process, we fed the $CNN$ with a set of intensity patterns measured in both planes at the fiber output and the related phase maps displayed on the SLM ($[x, |y|^2, |z|^2]$, using triplets of 64x64 images). The input field $x$ was sampled by the 34x34 actuators of the SLM to which we added some extra pixels to fill a 64x64 image. The amount of data (10 000 triplets) required to train the $CNN$ is similar to that needed for training the $TM$. However, the computing time is longer and lasted about 3h (NVIDIA RTX A1000 Laptop GPU), because of the

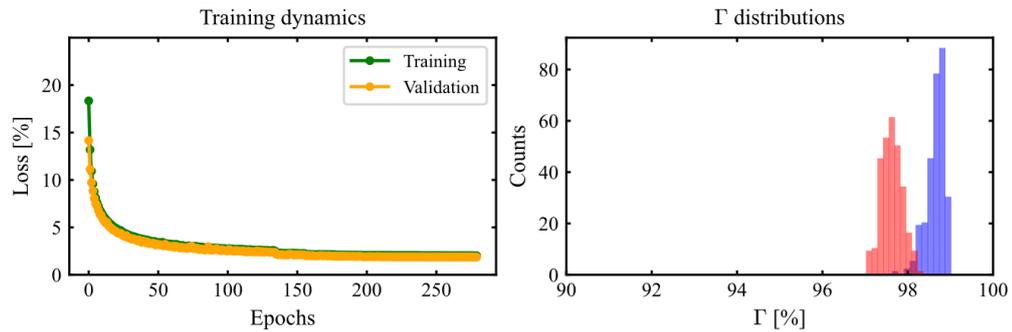

*Figure 10 : Left - CNN training dynamics – Right - Statistical distribution of the Pearson correlation coefficient between predicted and measured speckle patterns in the y (near field - blue) and z (far field - red) planes with the validation dataset.*

convolutional operations and the large number of 2D Fast Fourier Transform used in the loss function. The $CNN$ was optimized in less than 50 epochs (mini-batch size of 64 and learning rate 1e-3) and accurately predicted the near field $|y|^2$ and far field $|z|^2$ as shown Figure 10. The Pearson correlation

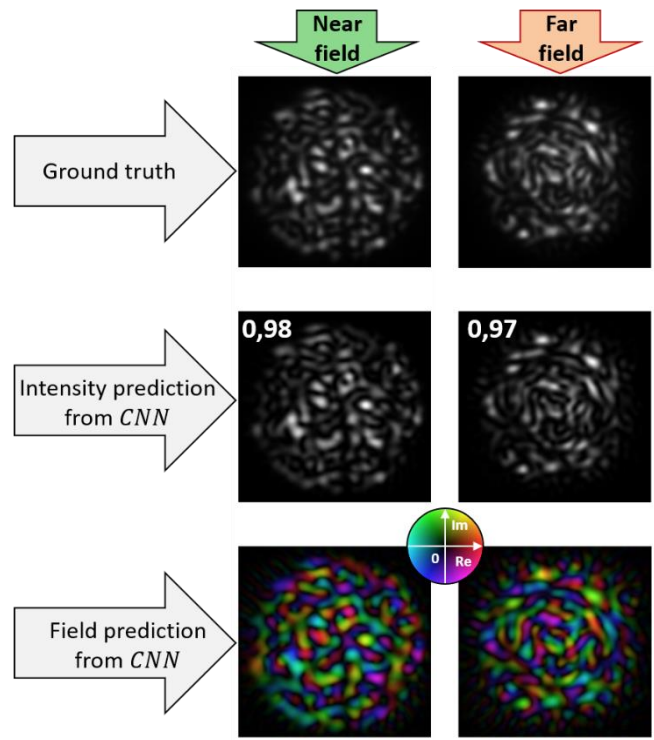

*Figure 11 : Example of intensity patterns obtained with a same input field x, in both y (left-hand column) and z (right-hand column) planes; Top row – experimental intensity patterns; middle row - corresponding intensity patterns predicted by CNN; Bottom row - corresponding complex fields predicted by CNN. Pearson correlation coefficients are indicated*

coefficients comparing ground truth (experimental patterns) and predicted intensity images in both near and far fields reach a very high level of respectively 98.5% and 97.5% (median values). Figure 11 shows with an example, the great similarity between the intensity images experimentally measured and those predicted by the $CNN$ in both plane for the same input field $x$.

As for the TM, we confirmed the good accuracy of the $CNN$ model with a second experiment by showing that the same model can be used to shape the MMF output beam in any plane, including planes different from those in which the learning data were measured. We optimized the input field $x$ (the SLM phase map $\phi$) with a gradient descent algorithm using the following loss function $\mathcal{L}_4$:

$$\min_{\phi \in [-\pi, +\pi[^n} \mathcal{L}_4 = 1 - \Gamma\left(\left|\mathcal{F}r_L\left(CNN(|x|e^{j\phi})\right)\right|, |z_L|\right) \qquad (8)$$

to produce a target pattern at a position $L$ from the fiber output. We conducted this optimization with the same $CNN$ for two image projection distances: $L = 0$ (MMF exit) and $L = 30\mu m$. The target shapes varied widely, encompassing symbols such as double dots, triangles, squares, crosses, spirals, and even words. Some experimental examples are shown in both planes Fig. 12. They are very similar to the predicted ones by the $CNN$, demonstrating its accurate ability to model the true complex

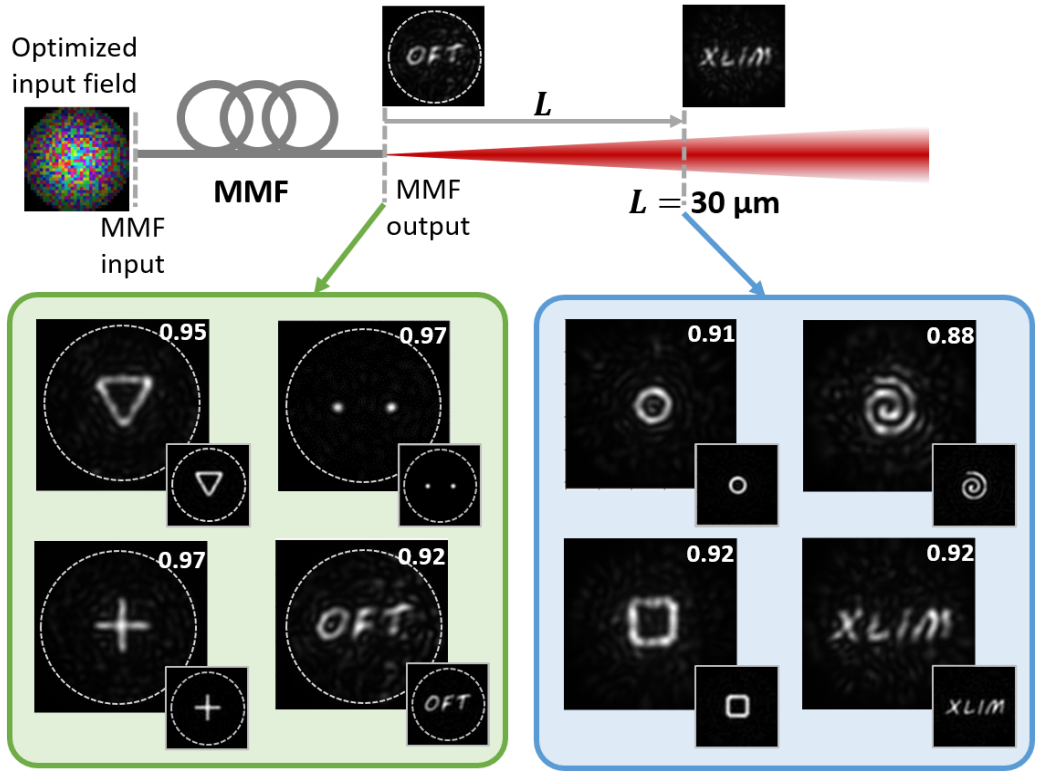

*Figure 12 : Examples of experimental shaped beams generated -Left – at the MMF output, -Right – 30μm away from the MMF output, -Insets - computed by the $CNN$; $\Gamma$ between computed and measured beams are indicated.*

transformation of the MMF.

In a prospective work, we explored our $CNN$ architecture to model a graded-index multimode fiber (GMMF) in a non-linear Kerr regime. In [33], U. Tegin et al. have investigated a recurrent neural network for prediction of spatiotemporal dynamics in GMMFs. In particular, their numerical work demonstrates that their NN can closely mimic the 2D spatial evolution along the fiber, in comparison with the outcomes of time-dependent beam-propagation method simulations. In our experimental study, we learned a $CNN$ model with conventional data (input field $x$ of random phase and corresponding output intensity $|y|^2$) aiming at predicting output scrambled intensities and we tested its generalization ability. In particular, we investigated self-cleaning on low order modes that was demonstrated in [34] using an optical feedback loop. In the setup of Fig. 1, we replaced the laser diode by an Nd:YVO$_4$ ultrafast laser, delivering 6.5ps pulses at 1064 nm and 1 MHz repetition rate (Sirius Spark Laser), with a peak power of 50kW. The GMMF of two-meter-long has a 26μm radius and a core-cladding index difference corresponding to a 0.21 numerical aperture. The fiber carries about 56 modes at 1064 nm per polarization. Compared to training the $CNN$ in the linear regime, we adapted the loss function to fit a relationship between the complex input fields illuminating the GMMF and their speckled patterns after propagation through the GMMF:

$$\mathcal{L}_{CNN_{NL}} = 1 - \Gamma(|CNN(x)|^2, |y|^2) \qquad (9)$$

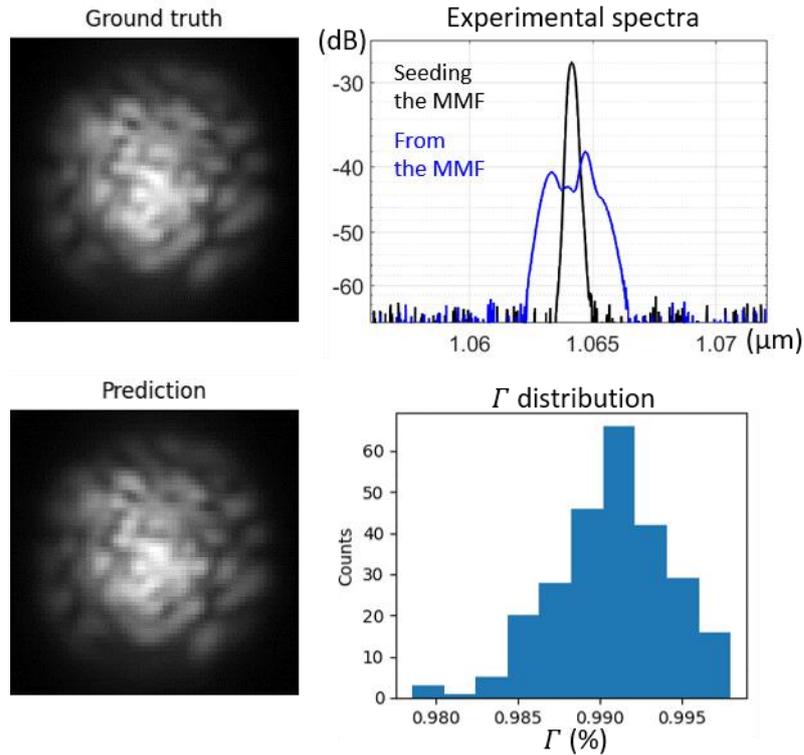

*Figure 14 Left – example of experimental (ground truth) and predicted intensity patterns. Top right – the corresponding experimental spectra (black: from the laser before seeding in the MMF – Blue: from the MMF. Bottom right: Pearson correlation distribution between the intensity predictions of the CNN and the experimental ground truth (validation set)*

Fig. 13 highlights the very good speckle predictions of the $CNN_{NL}$ with a median Pearson coefficient correlation of 99.2% between the ground truths and the intensities computed by the $CNN_{NL}$. These first results demonstrate the $CNN_{NL}$ ability to accurately mimic non-linear transformations. We further investigated beam self-cleaning by computing the input field using the previously employed

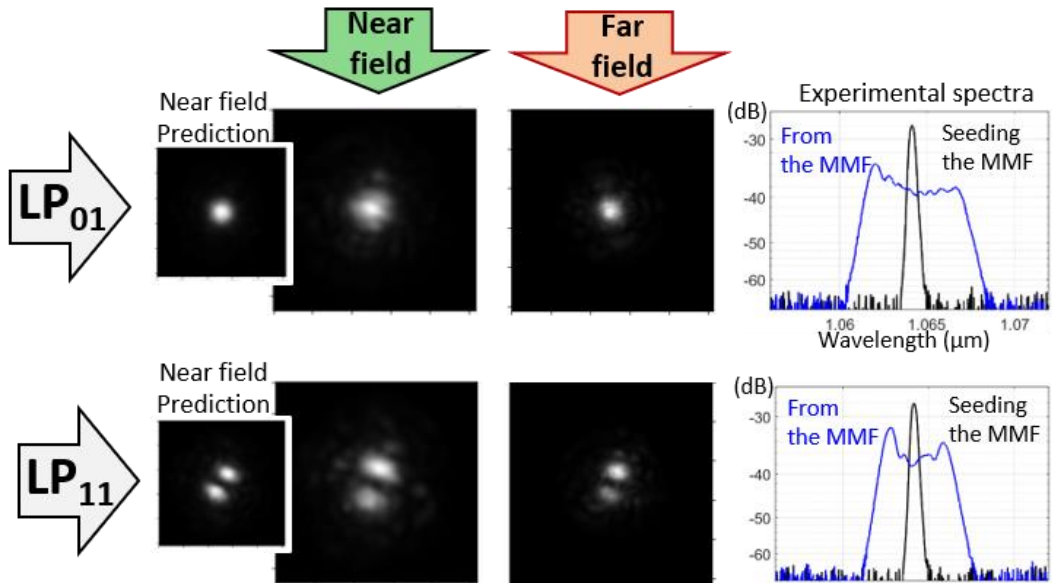

*Figure 13 : Examples of experimental LP modes generated by an input phase map optimized via the non-linear CNN -Top row: LP01 mode (experimental near field and far field) and the related spectrum density (Black: seeding the GMMF, Blue: from the GMMF), - Bottom row LP11 mode (experimental near field and far field) and the related spectrum density (Black: seeding the GMMF, Blue: from the GMMF), Insets: Target patterns computed by the non-linear CNN; $\Gamma$ between computed and experimental beams are indicated.*

optimization process (see equation 8 with $L = 0$). We determined the phase to be imprinted onto the

SLM, targeting a GMMF mode at the fiber exit that was not used during the training process. The computed phases for the LP$_{01}$ and LP$_{11}$ targets applied to the SLM generated the near and output fields of Fig. 14. The resulting images closely resemble to the targeted modes indicating beam cleaning was achieved despite this non-linear transformation not being explicitly learned. Notably, the non-linearity level involved during the training process was less than that in the shaping step. Specifically, the power coupled into the GMMF is lower for fine structuration of the SLM in case of speckle generation than for fostering low order mode emission (LP$_{01}$ and LP$_{11}$). This can be observed in Fig.13 and 14 by comparing the spectra widths after non-linear propagation in the GMMF. Thus, within the range of induced non-linearities in the experiments, our $CNN_{NL}$ was effective both in predicting output speckles and as part of beam shaping in the context of a Kerr self-cleaning. This is a proof of its generalization capability across various output beam profiles as well as various Kerr non-linearity levels.

5. Conclusion

We proposed a new method to retrieve full-complex models (Transmission Matrix and Neural Network) of a highly MMF (140 LP modes/polarization) with a machine learning approach, without the need of a reference beam. The models were optimized using triplets of data consisting of complex input fields, intensity images of the output fields, and corresponding far field patterns. The input field was structured by a segmented deformable mirror of 952 actuators. The full-complex models were first validated by the high similarity between the intensity images measured experimentally and predicted by the models at the fiber output in both near and far fields. The Pearson correlation coefficient reached 99.1% in the near field and 97.6% (median values) in the corresponding far field with the $TM$ and respectively 98.5% and 97.5% with the $CNN$. The accuracy of the models was also confirmed through their application in 3D beam shaping. This task is only feasible with a true full complex model. Using the retrieved $M$, we demonstrated its ability to find an input phase map that generates diverse arbitrary shapes ("digit" at 35µm from the MMF output and "letter" at 75µm), simultaneously in two output planes. With the $CNN$, we also demonstrated the possibility to shape the beam away from the MMF output even in planes where data had not been previously recorded. These abilities are not achievable with standard models (TM or NN) trained solely on a single plane with speckled patterns. There are some differences between the two models in terms of the amount of data needed for training and the computation time required. The main advantage of the $TM$ lies in its ability to characterize essential features of the MMF such as transmission channels via the SVD operation. It also requires less data to learn, particularly when dealing with a limited number of phase-modulation actuators. The $TM$ can also easily be inverted to compute the input field that generates any arbitrary output beam shape. In return, the MMF characteristics cannot be extracted from the $NN$. It requires more data to learn, but becomes competitive with the $TM$ training process as the number of actuators increases significantly. However, the $NN$ model is less sensitive to noise and non-linearity of the setup (SLM modulation, detection non-linearities). We also demonstrated that this architecture of $NN$ can model strong non-linearities such as Kerr effect. In contrast with the $TM$, the $NN$ also can accommodate minor perturbations in the setup, either by incorporating them during the training or ugh a short process of transfer learning. Depending on the intended application and the experimental conditions, one or the other of these true full-complex models can be selected.

**Funding.** This work was supported by the Région Nouvelle Aquitaine (AAPR2020-2019-8128410), the Agence AMIES (UAR3458) Labex and CPER: FEDER PILIM Nouvelle Aquitaine 2015-2027. For the purpose of Open Access, a CC-BY public copyright licence has been applied by the authors to the present document and will be applied to all subsequent versions up to the Author Accepted Manuscript arising from this submission.

**Acknowledgments.** The authors would like to thank the PLATINOM platform supported by the European Regional Development Fund and the Région Nouvelle Aquitaine (FEDER PILIM).

**Disclosures**. The authors declare no conflicts of interest.**References**

[1] D. J. Richardson, J. M. Fini, and L. E. Nelson, "Space-division multiplexing in optical fibres," Nat. Photonics 7, 354–362 (2013).
[2] B. J. Puttnam, et al., "Space-division multiplexing for optical fiber communications," Optica 8, 1186-1203 (2021).
[3] ] S. Rothe, et al., "Securing data in multimode fibers by exploiting mode-dependent light propagation effects". Research. 6:0065, DOI:10.34133/research.006 (2023)
[4] T. Čižmár and K. Dholakia, "Shaping the light transmission through a multimode optical fibre: complex transformation analysis and applications in biophotonics," Opt. Express 19, pp. 18871-18884 (2011).
[5] I.N. Papadopoulos, S. Farahi, C. Moser et al, "High-resolution, lensless endoscope based on digital scanning through a multimode optical fiber," Biomed. Opt. Express 4, 260–270 (2013).
[6] B. Lochocki, M. V. Verweg, J. J. M. Hoozeman et al., "Epi-fluorescence imaging of the human brain through a multimode fiber," APL Photonics 7, 7 071301 (2022).
[7] K. Wang, Y. Mizuno, X. Dong et al "Multimode optical fiber sensors: from conventional to machine learning-assisted," Meas. Sci. Technol. 35 022002 (2024)
[8] J. Montoya, C. Hwang, D. Martz, C. Aleshire, T. Y. Fan, and D. J. Ripin, "Photonic lantern kW-class fiber amplifier," Opt. Express 25, 27543-27550 (2017)
[9] R. Florentin, V. Kermene, J. Benoist, et al., "Shaping the light amplified in a multimode fiber," Light Sci. Appl. 6, e16208 (2017).
[10] H. Cao, T. Čižmár, S. Turtaev, et al. "Controlling light propagation in multimode fibers for imaging, spectroscopy, and beyond," Adv. Opt. Photon. 15, 524-612 (2023)
[11] Leite, I.T., Turtaev, S., Jiang, X. et al. Three-dimensional holographic optical manipulation through a high-numerical-aperture soft-glass multimode fibre. Nature Photon 12, 33–39 (2018)
[12] R. Florentin, V. Kermene, A. Desfarges-Berthelemot, A. Barthelemy, "Shaping of amplified beam from a highly multimode Yb-doped fiber using transmission matrix"Optics Express Vol. 27, Issue 22, pp. 32638-32648 (2019)
[13] Rothe, S.; Radner, H.; Koukourakis, N.; Czarske, J.W. Transmission Matrix Measurement of Multimode Optical Fibers by Mode-Selective Excitation Using One Spatial Light Modulator. Appl. Sci, 9, 195 (2019)
[14] D. Loterie, S. Farahi, I. Papadopoulos, A. Goy, D. Psaltis, and C. Moser, "Digital confocal microscopy through a multimode fiber," Opt. Express 23, 23845-23858 (2015)
[15] G. Konstantinou, A. Boniface, D. Loterie, E. Kakkava, D. Psaltis, C. Moser, "Improved two-photon polymerization through an optical fiber using coherent beam shaping," Optics and Lasers in Engineering, 160 (2023)
[16] J. Carpenter, B. C. Thomsen and T. D. Wilkinson, "Degenerate Mode-Group Division Multiplexing," in Journal of Lightwave Technology, vol. 30, no. 24, pp. 3946-3952, (2012)
[17] S. Cheng, T. Zhong and P. Lai, "Non-convex optimization for retrieving the complex transmission matrix of a multimode fiber," TENCON 2022 - 2022 IEEE Region 10 Conference (TENCON), Hong Kong, 2022, pp. 1-5.
[18] G. Huang, D. Wu, J. Luo, L. Lu, F. Li, Y. Shen, and Z. Li, "Generalizing the Gerchberg–Saxton algorithm for retrieving complex optical transmission matrices," Photon. Res. 9, 34-42, (2021)
[19] A. Drémeau, A. Liutkus, D. Martina, O. Katz, C. Schülke, F. Krzakala, S. Gigan, and L. Daudet, "Reference-less measurement of the transmission matrix of a highly scattering material using a DMD and phase retrieval techniques," Opt. Express 23, 11898-11911 (2015)
[20] G. Huang, D. Wu, J. Luo, Y. Huang, and Y. Shen, "Retrieving the optical transmission matrix of a multimode fiber using the extended Kalman filter," Opt. Express 28, 9487-9500 (2020)
[21] M. N'Gom, MB. Lien, N.M. Estakhri, et al. Controlling Light Transmission Through Highly Scattering Media Using Semi-Definite Programming as a Phase Retrieval Computation Method. Sci Rep 7, 2518 (2017)
[22] V. Tran, T. Wang, N. P. Nazirkar, et al "On the exploration of structured light transmission through a multimode fiber in a reference-less system". APL Photonics. 8 (12), 126111 (2023)
[23] S. Resisi, S. M. Popoff, Y. Bromberg, Image Transmission Through a Dynamically Perturbed Multimode Fiber by Deep Learning. Laser & Photonics Reviews (2021)
[24] U. Teğin, B. Rahmani, E. Kakkava, N. Borhani, C. Moser, and D. Psaltis, "Controlling spatiotemporal nonlinearities in multimode fibers with deep neuralnetworks," APL Photonics 5, 030804 (2020)